\documentclass[aps,prb,twocolumn,showpacs,superscriptaddress]{revtex4-1}

\usepackage{amssymb,amsmath}
\usepackage{graphicx}
\usepackage{dcolumn}
\usepackage{natbib}
\usepackage{longtable}
\usepackage{color}

\newcommand{\corr}[1]{{\color{black}{#1}}}

\begin{document}

\title{Spin-$\frac{1}{2}$ Kondo effect in an InAs nanowire quantum dot: the Unitary limit, conductance scaling and Zeeman splitting}

\author{Andrey V. Kretinin}
\email{andrey.kretinin@weizmann.ac.il}
\affiliation{Braun Center for Submicron Research, Condensed Matter Physics Department, Weizmann Institute of Science, Rehovot, Israel}
\author{Hadas Shtrikman}
\affiliation{Braun Center for Submicron Research, Condensed Matter Physics Department, Weizmann Institute of Science, Rehovot, Israel}
\author{David Goldhaber-Gordon}
\affiliation{Physics Department, Stanford University, Stanford, California, USA}
\affiliation{Braun Center for Submicron Research, Condensed Matter Physics Department, Weizmann Institute of Science, Rehovot, Israel}
\author{Markus Hanl}
\affiliation{Physics Department, Arnold Sommerfeld Center for Theoretical Physics, and Center for NanoScience,
Ludwig-Maximilians-Universit\"{a}t, Theresienstra{\ss}e 37, D-80333 M\"{u}nchen, Germany}
\author{Andreas Weichselbaum}
\affiliation{Physics Department, Arnold Sommerfeld Center for Theoretical Physics, and Center for NanoScience,
Ludwig-Maximilians-Universit\"{a}t, Theresienstra{\ss}e 37, D-80333 M\"{u}nchen, Germany}
\author{Jan von Delft}
\affiliation{Physics Department, Arnold Sommerfeld Center for Theoretical Physics, and Center for NanoScience,
Ludwig-Maximilians-Universit\"{a}t, Theresienstra{\ss}e 37, D-80333 M\"{u}nchen, Germany}
\author{Theo Costi}
\affiliation{Peter Gr\"{u}nberg Institut and Institute for Advanced Simulation,
Research Centre J\"{u}lich, 52425 J\"{u}lich, Germany}
\author{Diana Mahalu}
\affiliation{Braun Center for Submicron Research, Condensed Matter Physics Department, Weizmann Institute of Science, Rehovot, Israel}

\date{\today}

\begin{abstract}
We report on a comprehensive study of spin-$\frac{1}{2}$ Kondo effect in a strongly coupled quantum dot realized in a high-quality InAs nanowire. The nanowire quantum dot is relatively symmetrically coupled to its two leads, so the Kondo effect reaches the unitary limit. The measured Kondo conductance demonstrates scaling with temperature, Zeeman magnetic field, and out-of-equilibrium bias.
The suppression of the Kondo conductance with magnetic field is much stronger than would be expected based on a \textit{g}-factor extracted from Zeeman splitting of the Kondo peak. This may be related to strong spin-orbit coupling in InAs.

\end{abstract}


\maketitle

\section{Introduction \label{Introduction}}
The Kondo effect\cite{Kondo1964} is one of the most vivid manifestations of
many-body physics in condensed matter. First observed in 1930s in bulk metals through an anomalous increase in resistivity at low temperatures, it was later associated with the presence of a small amount of magnetic impurities.\cite{Sarachik1964} The modern theoretical understanding is that the single unpaired spin of the magnetic impurity forms a many-body state with conduction electrons of the host metal. This many-body state is characterized by a binding energy expressed as a Kondo temperature ($T_{\rm K}$). When the temperature is decreased below $T_{\rm K}$, the conduction electrons screen the magnetic impurity's unpaired spin, and the screening cloud increases the scattering cross-section of the impurity. More recently, advances in microfabrication opened a new class of experimental objects semiconductor quantum dots in which a few electrons are localized between two closely spaced tunneling barriers.\cite{Reed1988} At the same time, it had been theoretically predicted that an electron with unpaired spin localized in a quantum dot could be seen as an artificial magnetic impurity and, in combination with the electrons of the leads, would display the Kondo effect.\cite{Glazman1988,Ng1988} The first observation of Kondo effect in quantum dots was made in GaAs-based two-dimensional structures.\cite{Goldhaber-Gordon1998a,Goldhaber-Gordon1998,Cronenwett1998,Schmid1998,Simmel1999} Initially thought to be very difficult to observe in such experiments, the Kondo effect has now been seen in quantum dots based on a wide variety of nanomaterials such as
carbon nanotubes,\cite{Nygard2000,Quay2007} C$_{60}$ molecules,\cite{Yu2004,Scott2009} organic molecules,\cite{Li1998,Liang2002,Park2002,Yu2005} and semiconductor nanowires,\cite{Jespersen2006,Csonka2008,Nilsson2009,Kretinin2010} and has also been invoked to explain behavior of quantum point contacts.\cite{Cronenwett2002}

In this paper, we present a comprehensive study of the Kondo effect in a nanosystem of emerging interest, namely, InAs nanowires grown by the vapor-liquid-solid (VLS) method.\cite{Salfi2010} Building on initial reports of Kondo effect in InAs nanowires,\cite{Jespersen2006,Csonka2008} we report Kondo valleys with conductance near $2e^2/h$ in multiple devices and cooldowns. This high conductance, combined with temperature far below the Kondo temperature, allows quantitative measurements of conductance scaling as a function of temperature, bias, and magnetic field, which we compare to theoretical predictions independent of materials system. The high \textit{g}-factor and small device area, characteristic of InAs nanowires, allows measurement of the splitting of the zero-bias anomaly over a broad range of magnetic field, and we find that splitting is pronounced at lower magnetic field than predicted theoretically.

\section{Experiment}
The quantum dot from which data are presented in this paper is based on a 50~nm-diameter InAs nanowire suspended over a predefined groove in a $p^{+}$-Si/SiO$_{2}$ substrate and held in place by two Ni/Au (5nm/100nm) leads deposited on top of the nanowire. The leads' 450-nm separation defines the length of the quantum dot. The $p^{+}$-Si substrate works as a backgate. The InAs nanowire was extracted from a forest of nanowires grown by molecular beam epitaxy on a (011) InAs substrate using Au-catalyst droplets.
Wires from this ensemble were found to have a pure wurtzite structure, with at most one stacking fault per wire, generally located within 1~$\mu$m from the tip. We therefore formed devices from sections of nanowire farther from the wires' end, with a reasonable presumption that the active area of each device is free of stacking faults. Schottky barriers, and screening of the electric field from the gate electrode by the source and drain electrodes, together create potential barriers next to the metal contacts. Thus electrons must tunnel to the central part of the nanowire (the quantum dot) and the contacts, giving rise to Coulomb blockade (CB). An SEM image of a typical device is shown in Fig.~\ref{Fig1}(a). More details on growth, fabrication, and charging effects have been published previously.\cite{Kretinin2010}

Transport experiments were carried out in a dilution refrigerator with a base temperature $T_{base} \sim$ 10~mK. All experimental wiring was heavily filtered and thermally anchored to achieve electron temperature close to cryostat base temperature, as verified in shot noise measurements.\cite{Bid2010} Conductance measurements used standard lock-in techniques with a home-built ultra-low-noise transimpedance preamplifier operated at frequencies of $\sim$~2~kHz. Depending on the temperature $T$, the ac excitation bias was set in the range of 1-10 $\mu$Vrms to keep it equal to or smaller than $k_{B}T$ ($k_{B}$ is the Boltzmann constant). The magnetic field was applied perpendicular to both the substrate and the axis of the nanowire. A schematic representation of the nanowire-based device together with the experimental set-up is shown in Fig.~\ref{Fig1}(b).
\section{Results and discussion}
\begin{figure}
    \includegraphics[width=\columnwidth]{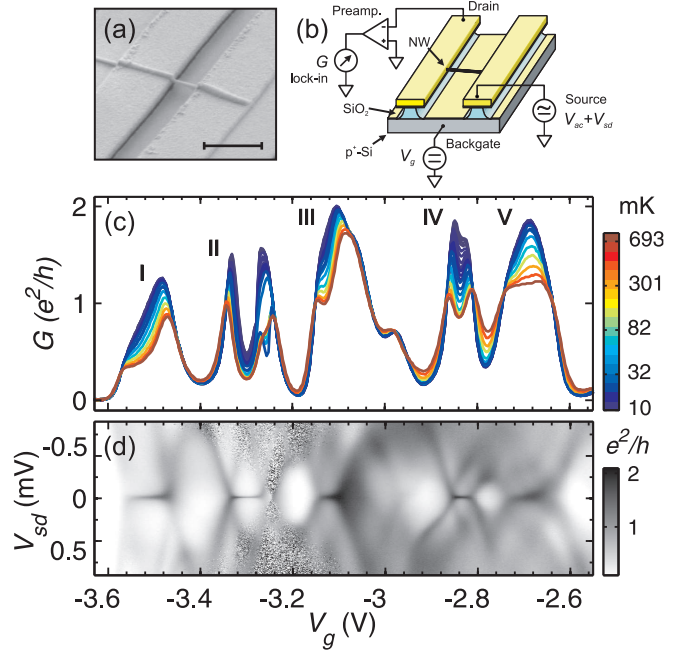}
        \caption{(a) SEM image of a typical suspended nanowire-based quantum dot device used in the experiment. The scale bar corresponds to 1 $\mu$m. (b) Schematic representation of the nanowire-based quantum dot device and its experimental setup. (c) The temperature dependence of the nanowire-based quantum dot conductance measured over a wide range of the backgate voltage $V_{g}$. Five Kondo valleys are labeled I through V here. This identification of valleys will be used throughout the paper. Discontinuities in the temperature-dependence in Valley II are caused by device instability at this particular range of $V_{g}$. (d) The gray-scale conductance plot in the $V_{g}-V_{sd}$ plane measured in the same range of $V_{g}$ as in (c) at temperature $T_{base} =$~10~mK. Panels (a) and (b) are adapted with permission from A. V. Kretinin \textit{et al.,} Nano Lett. \textbf{10}, 3439 (2010). Copyright $\copyright$ 2011 American Chemical Society.}
        \label{Fig1}
\end{figure}
First we would like to outline the main features associated with the Kondo effect which were studied in our experiment. The conductance of a quantum dot weakly coupled to leads is dominated by CB, seen as nearly periodic peaks in the conductance as a function of gate voltage, with the conductance strongly suppressed between peaks. Each peak signals a change in the dot occupancy by one electron. In contrast, a dot strongly coupled to the leads can show the Kondo effect, with the following signatures:\cite{Goldhaber-Gordon1998a,Cronenwett1998,Kogan2004} (1)~the Kondo effect enhances conductance between alternate pairs of Coulomb blockade peaks (that is, for odd dot occupancy). These ranges of enhanced conductance are conventionally termed ``Kondo valleys''. (2)~Conductance in Kondo valleys is suppressed by increasing temperature. (3)~Conductance in Kondo valleys is suppressed by applied source-drain bias ($V_{sd}$), giving rise to a zero-bias anomaly (ZBA). The full width at half maximum (FWHM) of the zero-bias peak is of the order of $4k_{B}T_{\rm K}/e$ ($e$ is the elementary charge). (4)~In contrast to the conductance in the CB regime whose upper limit is $e^{2}/h$,\cite{Grabert1992} the Kondo valley conductance can reach $2e^{2}/h$, equivalent to the conductance of a spin-degenerate 1D wire.\cite{Wiel2000} In this limit, ``valley'' is a misnomer, as the valley is higher than the surrounding peaks! (5)~The Kondo ZBA splits in magnetic field ($B$) with the distance between the peaks in bias being twice the Zeeman energy. (6) The dependence of the Kondo conductance on an external parameter $A$ such as temperature, bias or magnetic field can be calculated in the low- and high-energy limits.\cite{Pustilnik2004} In the low-energy limit, $k_{B}T_{\rm K} \gg A = \{k_{B}T,  eV_{sd},  |g|\mu_{B}B\}$, the conductance has a characteristic quadratic Fermi-liquid behavior:\cite{Nozi'eres1974,Schiller1995,Grobis2008,Scott2009}
\begin{equation}\label{QuadDep}
\textstyle
    G(A) = G_{0}\left[1-c_{A}\left(\frac{A}{k_{B}T_{\rm K}}\right)^{2}\right],
\end{equation}
where $G_{0} \equiv G(A=0)$
and $c_{A}$ is a coefficient of order unity. \corr{Its numerical value is different for each
parameter $A$, and depends on the definition of $T_{\rm K}$. In the present
paper, we use a convention\cite{Goldhaber-Gordon1998}
used in many experimental papers and define $T_{\rm K}$ by the relation
\begin{equation}
\label{eq:define-Tk}
G(T=T_{\rm K}) = 0.5 G_0 \; .
\end{equation}
}
In the opposite limit of high energy, when $k_{\rm B}T_{\rm K} \ll A$, the conductance shows a logarithmic dependence. For example, as a function of temperature:\cite{Kondo1964,Ng1988}
\begin{equation}\label{LogDep}
\textstyle
    G(T) \propto G_{0}/\ln^{2}\left(\frac{T}{T_{\rm K}}\right).
\end{equation}
There is no analytical expression for the intermediate regime, where the parameter $A \approx k_{B}T_{\rm K}$, but Numerical Renormalization Group (NRG) calculations~\cite{Costi1994} show that the connection between one limit and the other is smooth and monotonic, without any sharp feature at $A=k_{B}T_{\rm K}$.

Before detailed consideration and discussion of the results we give a broad overview of the experimental data used in this study. It will be followed by three subsections focusing on the observed unitary limit of the Kondo effect (Sec.~\ref{Kondo effect in the Unitary limit}), conductance scaling with different external parameters (Sec.~\ref{Universal scaling}), and some peculiarities observed in the Zeeman splitting (Sec.~\ref{Zeeman splitting}).

Figure~\ref{Fig1}(c) presents the linear conductance $G$ as a function of the backgate voltage $V_{g}$. Different color corresponds to different temperature, ranging from 10~mK to 693~mK. The Kondo effect modifies the CB peaks so strongly that the separate peaks are no longer recognizable and the simplest way to identify Kondo valleys is to look at the the gray-scale plot of differential conductance as a function of both $V_{g}$ and $V_{sd}$ (``diamond plot''), Fig.~\ref{Fig1}(d). Every Kondo valley is marked by a ZBA seen as a short horizontal line at $V_{sd} =0$. Different widths of ZBAs on the gray-scale plot reflect differences in the Kondo temperature. In these same Kondo valleys, conductance decreases with increasing temperature (Fig.~\ref{Fig1}(c)). \corr{Note that Kondo valleys alternate with valleys having opposite temperature dependence or almost no temperature-dependence, corresponding to even occupancy of the quantum dot. A small unnumbered peak at about $V_{g}=$~-2.95~V departs from the general pattern of conductance observed in the experiment. Most likely, this feature, which occurs for even occupancy, is associated with transition to a triplet ground state, and thus emergence of spin-1 and singlet-triplet Kondo effect.\cite{Kogan2003,Granger2005,Roch2008} However, it is difficult to conclusively identify the nature of this anomaly since its temperature and bias dependencies are weak.}

All conductance peaks shown in Fig.~\ref{Fig1}(c) exceed $e^2/h$, reflecting Kondo-enhanced conductance and relatively symmetric coupling to the two leads. In particular, conductance around $V_{g} =$~-3.1~V in Valley III reaches the unitary limit of $2e^{2}/h$, to within our experimental accuracy.
\subsection{Kondo effect in the unitary limit \label{Kondo effect in the Unitary limit}}
\begin{figure}
  \includegraphics[width=\columnwidth]{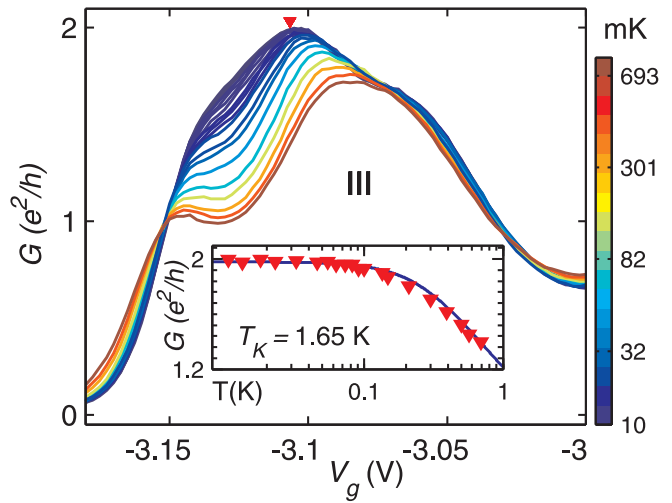}
  \caption{The Kondo effect in its unitary limit. The main plot shows the linear conductance $G$ in Valley III, as a function of backgate voltage $V_{g}$ at different temperatures. The dark blue curve corresponds to the lowest temperature of 10~mK. Inset: The red triangles correspond to the temperature dependence of the conductance at a fixed $V_{g} =$~-3.107~V (marked by the red triangle in the main graph). The blue curve represents the result of approximation with Eq.(\ref{TempDep}) where $G_{0} = 1.98 e^{2}/h$ and $T_{\rm K} =$~1.65~K.}
  \label{Fig2}
\end{figure}
To realize maximum conductance in resonant tunneling, the quantum dot should be symmetrically coupled to the leads. In the conventional case of CB, electrostatic charging allows only one spin at a time to tunnel, limiting the maximum conductance through the dot to $e^{2}/h$.\cite{Grabert1992} The Kondo effect dramatically changes the situation by forming a spin-degenerate many-body singlet state, enabling both spins to participate in transport in parallel so that Kondo conductance can reach its unitary limit at $2e^{2}/h$.\cite{Glazman1988,Ng1988} Experimentally, the unitary limit, first observed by van der Wiel \textit{et al.}\cite{Wiel2000} in a GaAs-based gate-defined quantum dot, remains the exception rather than the rule, because it requires being far below the Kondo temperature, having symmetric tunnel coupling to the two leads, and having precisely integer dot occupancy.

Figure~\ref{Fig2} presents a zoomed-in view of Valley III from Fig.~\ref{Fig1}(c), showing the Kondo effect in the unitary limit. Note how the conductance maximum gradually approaches $2e^{2}/h$ with decreasing temperature. Here the limit is reached only at some particular $V_{g}$, showing a peak instead of an extended plateau as reported by van der Wiel \textit{et al.}.\cite{Wiel2000}
Since tunneling is so strong that level widths are almost as large as the Coulomb interaction on the dot, the dot occupancy $n_{d}$ is not well-quantized but rather changes monotonically, passing through $n_{d}=1\ (n_\uparrow=n_\downarrow=1/2)$ at $V_g \approx -3.1$V, where the unitary limit is observed. In accordance with the Friedel sum rule the conductance of the dot is predicted to depend on the dot occupancy $n_{\uparrow,\downarrow}$ as $G(\uparrow,\downarrow)=(e^{2}/h)\sin^{2}(\pi n_{\uparrow,\downarrow})$. So the sum of the conductances is $2e^{2}/h$ when $n_{d}=1$. Note that the Kondo conductance shown in Fig.~\ref{Fig1}(c) always exceeds 1.3~$e^{2}/h$ for different dot occupancies, showing that the wave function overlap with the two leads is rather equal: the two couplings are within a factor of four of each other over this whole range, suggesting that disorder along the nanowire and especially at the tunnel barriers is quite weak. To extract the Kondo temperature we apply a widely used phenomenological expression\cite{Goldhaber-Gordon1998a} for the conductance $G$ as a function of temperature:
\begin{equation}
\textstyle
    G(T) = G_{0}\left[1+\left(T/T'_{K}\right)^{2}\right]^{-s},
    \label{TempDep}
\end{equation}
where $G_{0}$ is the zero-temperature conductance, $T'_{K} = T_{\rm K}/(2^{1/s} - 1)^{1/2}$, and the parameter $s =$~0.22 was found to give the best approximation to NRG calculations for a spin-$\frac{1}{2}$ Kondo system.\cite{Costi1994} Here, the definition of $T_{\rm K}$ is such that $G(T_{\rm K})=G_{0}/2$. The inset of Fig.~\ref{Fig2} shows the conductance for different temperatures at $V_{g} = $~-3.107~V (marked by the red triangle in the main figure). The blue curve in the inset represents the result of the data approximation using Eq.(\ref{TempDep}) where the fitting parameters $G_{0}$ and $T_{\rm K}$ are $(1.98\pm0.02)e^{2}/h$ and 1.65$\pm$0.03~K,\footnote{Here and throughout the text the errors and error bars represent the 68\% confidence interval. The systematic errors caused by the uncertainty in conductance and temperature measurements are believed to be less than 3\%} respectively, showing that the system is in the ``zero-temperature'' limit at base temperature, $T_{\rm K}/T_{\rm base} \approx 165$.

\subsection{Conductance scaling with temperature, magnetic field and bias\label{Universal scaling}}
As noted above, the Kondo conductance as a function of temperature, bias or magnetic field should be describable by three universal functions common for any system exhibiting the Kondo effect. Before discussing expectations for universal scaling we describe in detail how
temperature, magnetic field and bias affect the Kondo conductance in our experimental system.
\subsubsection{Kondo conductance and Kondo temperature at zero magnetic field\label{Kondo at zero field}}
\begin{figure}
  \includegraphics[width=\columnwidth]{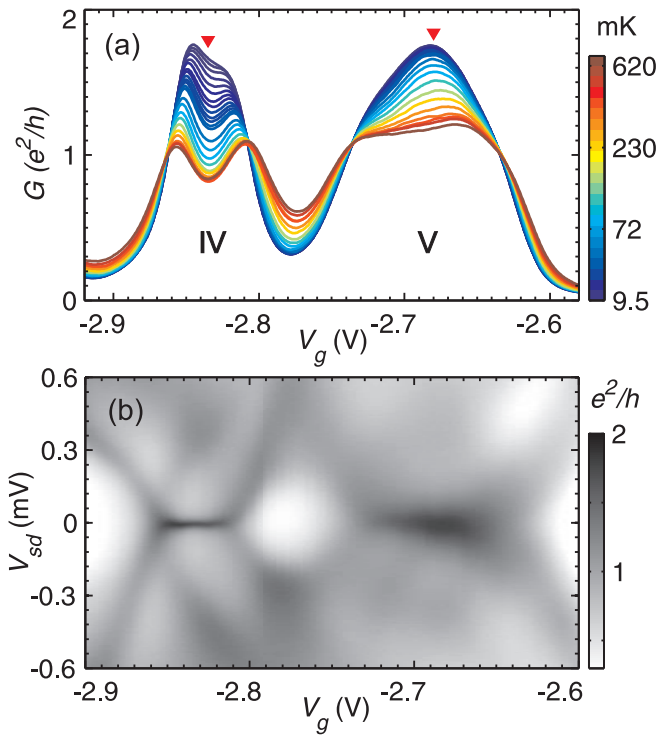}
  \caption{(a) The detailed measurement of the conductance temperature dependence shown in Fig.~\ref{Fig1}(c), Valleys IV and V. The red triangles mark two values of $V_{g} =$~-2.835~V and $V_{g} =$~-2.680~V for which the conductance as a function of $V_{sd}$ is plotted in Fig.~\ref{Fig4}(a) and Fig.~\ref{Fig4}(b), respectively. (b) The gray-scale conductance plot in the $V_{g}-V_{sd}$ plane was measured in the same range of $V_{g}$ as in (a), at temperature $T =$~10~mK.}
  \label{Fig3}
\end{figure}
 \begin{figure}[!ht]
  \includegraphics[width=\columnwidth]{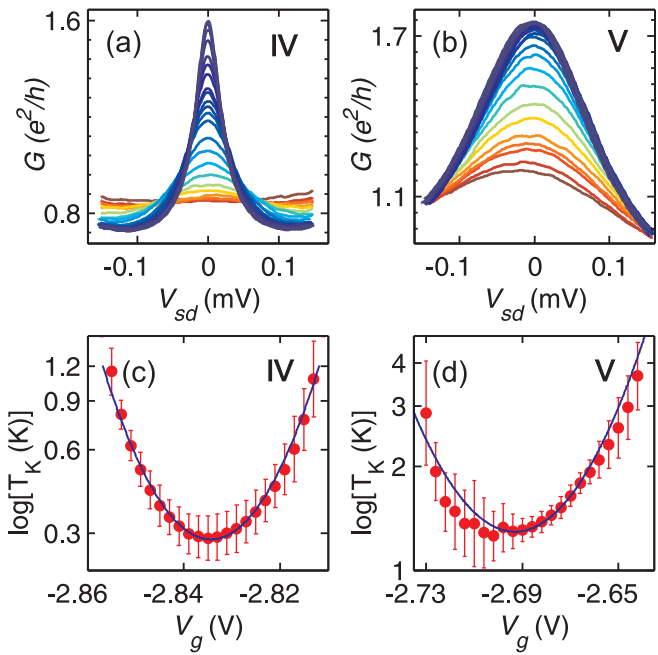}
  \caption{Nonlinear conductance as a function of $V_{sd}$ around zero bias for different temperatures at  $V_{g} =$~-2.835~V (a) and $V_{g} =$~-2.680~V (b), near the centers of Kondo valleys IV and V. The color scale is as in Fig.~\ref{Fig3}(a).  (c,d) The Kondo temperature $T_{\rm K}$, plotted on a semi-log scale, as a function of $V_{g}$ for these same valleys. Panel (c) corresponds to Valley IV and panel (d) to Valley V. Blue curves in both panels show fits of Eq.~(\ref{KondoTemp}) to data, with $\Gamma_{IV}\approx$~176$\mu eV$  for Valley IV and $\Gamma_{V}\approx$~435$\mu eV$ for Valley V.}
  \label{Fig4}
\end{figure}
For a more detailed look at the spin-$\frac{1}{2}$ Kondo effect at $B = 0$, we select the two Kondo valleys IV and V (see Fig.~\ref{Fig1}(c)). The zoomed-in plot of these two valleys is shown in Fig.~\ref{Fig3}(a,b). The coupling to the leads, and hence the Kondo temperature, is much larger in Valley V than in Valley IV. Valley IV shows a typical example of how two wide Coulomb blockade peaks merge into one Kondo valley as the temperature decreases below $T_{\rm K}$.\cite{Goldhaber-Gordon1998,Cronenwett1998,Wiel2000} Valley V, in contrast, does not evolve into separate CB peaks even at our highest measurement temperature of 620~mK. Also, as seen from Fig.~\ref{Fig3}(b), the width of the ZBA, which is proportional to $T_{\rm K}$, is larger for Valley V. To illustrate this, in Fig.~\ref{Fig4}(a,b), we plot the conductance as a function of $V_{sd}$ at different temperatures for two values of $V_{g}$ (marked by red triangles in Fig.~\ref{Fig3}(a)) corresponding to the two valleys. In addition to the ZBA of Valley IV being significantly narrower than that of Valley V, at the highest temperatures, the ZBA of Valley IV is completely absent, while the ZBA of Valley V is still visible, pointing to a significant difference in $T_{\rm K}$. To quantify this observation we found $T_{\rm K}$ as a function of $V_{g}$ for both valleys by fitting the temperature-dependent conductance using Eq.~(\ref{TempDep}). The result of this fit is presented in Fig.~\ref{Fig4}(c,d). $T_K$ shows a parabolic evolution across each Valley, with $T_{\rm K}$ ranging from 0.3~K to 1~K for Valley IV and from 1.3~K to 3~K for Valley V. This significant difference in $T_{\rm K}$ correlates with the difference in the ZBA width shown in Fig.~\ref{Fig4}(a,b). However, the relation between the FWHM of the ZBA peak and $T_{\rm K}$ is more ambiguous due to out-of-equilibrium physics.\cite{Anders2008}

To understand the dependence of $T_{\rm K}$ on $V_{g}$ and to extract
some relevant parameters of the system we use \corr{an analytic}
prediction for the dependence of the Kondo temperature \corr{based on the}
microscopic parameters in the Kondo regime of the single-impurity
Anderson model: \cite{Haldane1978}
\begin{equation}
    T_{\rm K} = \corr{\eta_{_\mathrm{NRG}}} \cdot \tfrac{\sqrt{\Gamma U}}{2} \exp\left[
       \tfrac{\pi\varepsilon_{0}(\varepsilon_{0} + U)}{\Gamma U}
    \right]
\text{.}\label{KondoTemp}%
\end{equation}
Here $\Gamma$ is the width of the resonant tunneling peak, $U =
e^{2}/C_{tot}$ is the charging energy ($C_{tot}$ is the total
capacitance of the dot), and $\varepsilon_{0}$ is the energy of the
resonant level relative to the Fermi level. As $T_{\rm K}$ is derived from the conductance [c.f. text
following Eq.~(\ref{TempDep})], the prefactor $\eta_{_\mathrm{NRG}}$
in Eq.~(\ref{KondoTemp}) of order unity was calibrated using the NRG.
To this end, we calculated the conductance $G(T)$ for the
single-impurity Anderson model at $\varepsilon_0 = -U/2$, for fixed
$U/\Gamma \simeq 4.5$. The requirement that $G(T=T_{\rm K})/G(0) = G_0/2$ fixes
the prefactor in $T_{\rm K}$ to \corr{$\eta_{_\mathrm{NRG}} \simeq
1.10$, which we took constant throughout.
$\eta_{_\mathrm{NRG}}$ does vary as a function
of $U/\Gamma$ within a few tens of percent, due to the exponential sensitivity of
Eq.~(\ref{KondoTemp}), however, since $U$ and $\Gamma$ are already pretty well-constrained
in our case, this results in negligible variations in our fitted $U$, $\varepsilon_0$ or
$\Gamma$.}

To determine the parameters $U$, $\varepsilon_0$ and $\Gamma$, we proceed as follows. The value of $U\approx 400 \mu$eV was found from Fig.~\ref{Fig3}(b) for Valley IV (we assume the value is equal for Valley V, though it may be slightly lower, given the stronger tunnel coupling there). To relate $\varepsilon_{0}$ and $V_{g}$, we used a simple linear relation $V_{g} - V_{g0} = \alpha\varepsilon_{0}$ with the lever arm $\alpha = C_{tot}/C_{g}$, where $V_{g0}$ is the position of the Coulomb peak and $C_{g}$ is the gate capacitance. Here $C_{tot}=e^{2}/U$ and $C_{g} = e/\Delta V_{g}$ where  $\Delta V_{g}$ is the CB period. $\Gamma$ was determined by fitting the curvature of $\log T_K$ with respect to gate voltage in Fig.~\ref{Fig4}(c,d), yielding $\Gamma_{IV}\approx$~176~$\mu$eV and $\Gamma_{V}\approx$~435~$\mu$eV for Valley IV and V, respectively.

As noted above, the predicted dependence of $T_{\rm K}$ in Eq.~(\ref{KondoTemp}) is based on the Anderson model in the Kondo regime ($\varepsilon_{0}/\Gamma < -1/2$).\cite{Haldane1978} The fitting of the data with Eq.~(\ref{KondoTemp}), however, gave $\varepsilon_{0}/\Gamma_{IV}\sim -1.1$ and $\varepsilon_{0}/\Gamma_{V} \sim -0.5$ in the centers of Valley IV and V, respectively. So the Kondo regime $\{|\varepsilon_{0}|,|\varepsilon_{0}+U|\} > \Gamma/2$ is reached only near the center of Valley IV and only at the very center of Valley V. The rest of the gate voltage range in these Valleys is the mixed valence regime, where charge fluctuations are important and Kondo scaling should not be quantitatively accurate.\cite{Makarovski2007} Note: our NRG calculations show that the deviations from universal scaling up to $\varepsilon_{0}\sim-\Gamma/2$ should be small for $T<T_K$. In any case, we have not attempted to take into account multiple levels in our calculations, which could quantitatively but not qualitatively modify the predicted behaviors.

\subsubsection{Kondo conductance at non-zero magnetic field\label{Kondo at non-zero field}}
The Kondo effect in quantum dots at non-zero magnetic field is predicted and observed to exhibit a Zeeman splitting of the ZBA by an energy $\Delta = 2|g|\mu_{\rm B}B$\cite{Goldhaber-Gordon1998a,Cronenwett1998} ($g$ is the $g$-factor, $\mu_{\rm B}$ is the Bohr magneton), which is a direct consequence of the (now-broken) spin-degeneracy of the many-body Kondo singlet.\cite{Meir1993,Costi2000}

\begin{figure}
  \includegraphics[width=1.00\columnwidth]{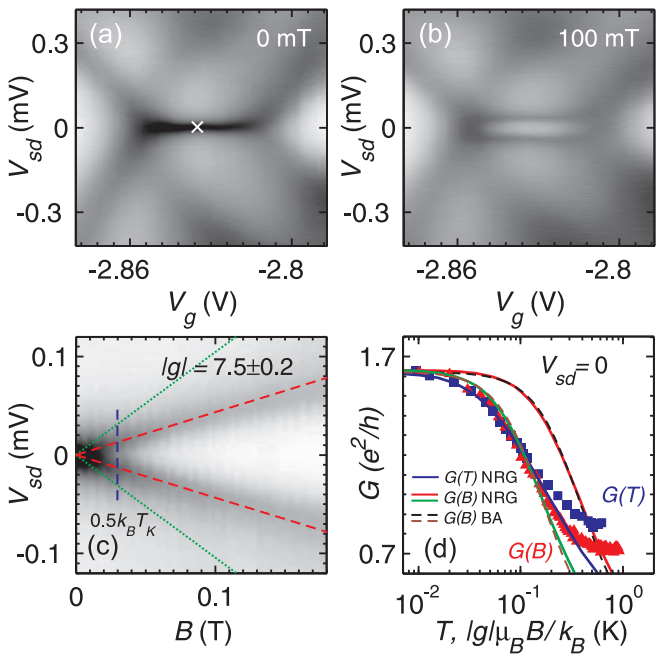}
  \caption{The Zeeman splitting of the Kondo ZBA measured at $T=$~10~mK. (a) The gray-scale conductance  plot of Kondo valley IV (see Fig.~\ref{Fig3}(a)) measured at $B=$~0. (b) The same as in (a) but at $B=$~100~mT. (c) Gray-scale conductance plot in the $V_{sd}-B$ plane measured at fixed $V_{g}=$~-2.835~V denoted by the cross in panel (a). The red dashed lines represent the result of the fitting with expression $V_{sd} = \pm|g|\mu_{\rm B}B/e$, where $|g|=$~7.5$\pm$0.2. Vertical blue dashed line marks magnetic field value 0.5$k_{B}T_{\rm K}/|g|\mu_{\rm B}$ as a reference for the onset of Zeeman splitting (here $T_{\rm K}=$~300~mK). While $|g|=7.5$ gives the best match to linear Zeeman splitting, $|g|=18$ (green dotted lines) could account for the fact that Zeeman splitting is resolved at very low field. (d) Conductance at $V_{sd}=$~0 as a function of $T$ (blue squares) and as a function of the effective temperature $T_{B} \equiv |g|\mu_{\rm B}B/k_{\rm B}$ (red triangles). The solid blue curve shows $G(T)$ from NRG, the solid red curve $G(B)$ from NRG, and the dashed black curve $G(B)$ from exact Bethe Ansatz (BA) calculations for the Kondo model.\cite{Costi2001,Andrei1982} \corr{These assume $|g|=7.5$. For NRG and BA calculations of magnetic field dependence, additional curves (solid green and dashed brown) are plotted for $|g|=18$, showing better match to linear conductance data -- though not to the differential conductance in (c) above.}}
  \label{Fig5}
\end{figure}

To analyze the Zeeman splitting in our nanowire-based quantum dot we focus on Kondo Valley IV. The Kondo ZBA at zero field, seen in a zoom-in in Fig.~\ref{Fig5}(a), is suppressed at $B = 100$~mT, but recovers once a bias of $\sim$~40~$\mu V$ is applied (Fig.~\ref{Fig5}(b)). Contrary to earlier observations in InAs nanowires,\cite{Csonka2008} we find that the $g$-factor at a given field is independent of $V_{g}$ as illustrated by the parallel slit-like shape of the Zeeman splitting (Fig.~\ref{Fig5}(b)).\footnote{The \textit{g}-factor measured for Valley III at $V_{g} =$~-3.12 V is $|g| =$~7.5$\pm$0.2. Unfortunately, it was problematic to extract the \textit{g}-factor reliably for Valley V due to large $\Gamma_{V}$ and it was hence assumed to be the same as for Valley IV. The \textit{g}-factor for Valley I measured at $V_{g} =$~-3.5 V (see Fig.~\ref{Fig1}(c)) turns out to be somewhat larger $|g| =$~8.7$\pm$0.2.} The gray-scale conductance plot in Fig.~\ref{Fig5}(c) presents the evolution of the Zeeman splitting with magnetic field at fixed $V_{g} =$~-2.835~V, marked by the cross in Fig.~\ref{Fig5}(a) (for the associated ZBA measured at $B =$~0 refer to Fig.~\ref{Fig4}(a)). The plot shows the splitting in bias $\Delta/e$ to be almost linear in magnetic field, which allows us to deduce the value of the $g$-factor by fitting the data with a linear dependence $V_{sd} = \pm|g|\mu_{\rm B}B/e$ for $30$~mT$ < B < 100$~mT. Two red lines in Fig.~\ref{Fig5}(c) show the result of fitting with $|g|=$~7.5$\pm$0.2 (the meaning of the dotted green lines will be discussed below). This number is smaller by a factor of two than the InAs bulk value of $|g|=15$, possibly due to the reduced dimensionality of the nanowire device,\cite{Bjork2005} and it is consistent with previous measurements.\cite{Jespersen2006}

We now compare the dependence of the Kondo conductance on the temperature and magnetic field, respectively. In order to do so we plot on the same graph $G(T,B=0)$ and $G(T=T_{\rm base}, B)$ both taken in equilibrium at $V_{g}=$~-2.835~V (Fig.~\ref{Fig5}(d)). In order to quantitatively compare the effect of magnetic field to that of temperature, we associate each magnetic field value with an effective temperature $T_{B}(B) \equiv |g|\mu_{\rm B}B/k_{\rm B}$, where $|g|=7.5$ is extracted from the linear Zeeman splitting of peaks in differential conductance. The comparison of the linear conductance data is presented in Fig.~\ref{Fig5}(d), where $G(T)$ is shown by the blue squares, $G(B)$ by the red triangles. In this same plot, theoretical predictions are shown as curves: blue for $G(T)$ and red for $G(B)$. Note that for $|g|=7.5$ (this value extracted from the splitting of the differential conductance peaks), the blue and red curves differ substantially for essentially all nonzero values of their arguments, with magnetic field having a much weaker predicted effect than temperature. Surprisingly, in light of this theoretical prediction, the two sets of experimental data lie almost on top of one another up to about 200~mK~$\approx T_{\rm K}$.
The NRG results for $G(T=0, B)$ \cite{Costi2000,Costi2001} have been checked against exact Bethe ansatz calculations \cite{Andrei1982,Costi2000} for $G(T=0,B)$ (dashed black curve in Fig.~\ref{Fig5}(d)) and are seen to be in excellent agreement, \corr{so the disagreement between theory and experiment is not related to a particular calculational framework. Were we to assume $|g|=18$, we could explain the experimental magnetic field dependence of linear conductance $G(T=0, B)$, as shown by alternative curves (solid green and dashed brown) plotted in ~\ref{Fig5}(d). This value of $g$ is within the realm of possibility for InAs nanowires.\cite{Csonka2008} However, we are inclined to rely on the $g$ value of 7.5 extracted from the splitting of the peaks in the differential conductance. With $|g|=18$ we would have the puzzling result that the splitting of peaks in differential conductance would be less than half the expected $2|g|\mu_B B$ (see dotted green lines in Fig.~\ref{Fig5}(c)), which would be hard to explain. Regardless, the mismatch between the strength of magnetic field effects on linear and differential conductance is a conundrum. We hope this work will stimulate further theory and experiment to address this issue.}

\subsubsection{Universal conductance scaling\label{Scaling subsection}}

\begin{figure}
  \includegraphics[width=\columnwidth]{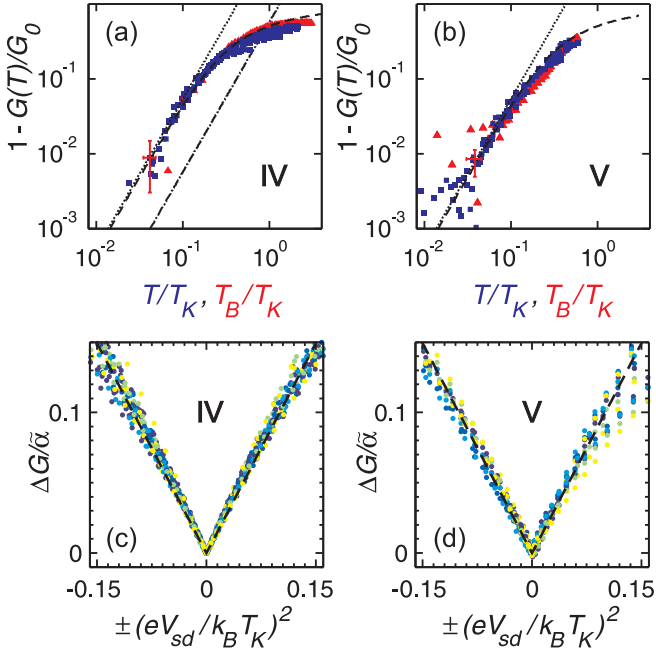}
  \caption{(a,b) The equilibrium conductance of Kondo valleys IV (a) and V (b) at different $V_{g}$, scaled as a function of a single argument $T/T_{\rm K}$  (blue squares) and $T_{B}/T_{\rm K}$  (red triangles), where $T_{B} \equiv |g|\mu_{\rm B}B/k_{\rm B}$. The dashed curve shows the universal function described by Eq.~(\ref{TempDep}). The dotted line represents the low-energy limit of Eq.~(\ref{QuadDep}) with $c_{A} = c_{T} =$~5.38. The dash-dotted line shows the theoretically predicted low-field scaling of $G(B)$ with $c_{B} =$~0.55. The values of $G_{0}$ and $T_{\rm K}$ were found by fitting the data with Eq.~(\ref{TempDep}), see Sec.~\ref{Kondo at zero field}. For values of $V_{g}$ refer to Fig.~\ref{Fig4}(c,d) and Fig.~\ref{Fig5}(d). (c,d) The scaled conductance $\Delta G/\tilde{\alpha}=(1-G(T,V_{sd})/G(T,0))/\tilde{\alpha}$, where $\tilde{\alpha} = c_{T}\alpha/(1+c_{T}(\gamma/\alpha-1))(T/T_{\rm K})^{2}$, versus $(eV_{sd}/k_{\rm B}T_{\rm K})^{2}$ taken at several $V_{g}$ along Kondo valleys IV (c) and V (d). For valley IV the backgate voltage was chosen from the range $V_{g}=$~-2.82~V to -2.85~V with 5~mV step and for valley V from the range $V_{g}=$~-2.68~V to -2.72~V with 20~mV step. Different colors of the data points represent different temperatures (9.5~mK, 12.9~mK, 22.4~mK, 32.6~mK, 46.1~mK, 54.2~mK). The dashed line shows the corresponding scaling function given by Eq.~(\ref{VsdScaling}) with $\alpha=$~0.18 and $\gamma=$~1.65.}
  \label{Fig6}
\end{figure}

In testing universal conductance scaling, we concentrate first on the scaling of the linear conductance with $T$ and $B$. In the case of temperature dependence, the universal scaling function has the form of Eq.~(\ref{TempDep}). This expression has been applied to a wide variety of experimental Kondo systems\cite{Goldhaber-Gordon1998,Nygard2000,Jespersen2006,Scott2009} and after expansion in the low-energy limit ($T/T_{\rm K}\ll1$) it becomes Eq.~(\ref{QuadDep}) describing the quadratic dependence on temperature:\cite{Grobis2008}
\begin{equation}\label{TempQuad}
\textstyle
    G\approx G_{0}\left[1-c_{T}\left(T/T_{\rm K}\right)^{2}\right],
\end{equation}
where $c_{T}=c_{A}=s(2^{1/s}-1)=$~4.92 and $s=$~0.22 is taken from Eq.~(\ref{TempDep}).
Note that this coefficient $c_{T}$ is about 10\% smaller than the more reliable value $c_{T} \approx$~5.38\cite{Nozi'eres1974,Costi1994,Konik2001,Konik2002} found from the NRG calculations on which the phenomenological form of Eq.~(\ref{TempDep}) is based (see Table~\ref{TableI}).\footnote{This slight disagreement stems from the fact that the phenomenological expression given by Eq.~(\ref {TempDep}) was designed for the intermediate range of temperatures and does not necessarily describe the dependence accurately at asymptotically low $T\ll T_{\rm K}$ or asymptotically high $T\gg T_{\rm K}$ temperatures. Hereafter, for the low-temperature analysis we use the theoretically predicted value $c_{T} =$~5.38} Since Eq.~(\ref{TempDep}) is independent of the particular system, it can be used as the universal scaling function $G/G_{0}=f(T/T_{\rm K})$. Figures~\ref{Fig6}(a,b) show the equilibrium Kondo conductance $(1-G/G_{0})$ of valleys IV and V (see Fig.~\ref{Fig3}(a)) plotted as a function of $T/T_{\rm K}$, taken at different $V_{g}$. Here, the values of $G_{0}$ and $T_{\rm K}$ are found by fitting the data with Eq.~(\ref{TempDep}) for $T\leq 200$mK (for higher temperatures the conductance starts to deviate from the expected dependence due to additional high-temperature transport mechanisms).
As seen in Fig.~\ref{Fig6}(a,b), all the data collapse onto the same theoretical curve (dashed) regardless of the values of $V_{g}$ or $T_{\rm K}$. In the low-energy limit $T/T_{\rm K}<$~0.1 the conductance follows a quadratic dependence set by Eq.~(\ref{QuadDep}) with coefficient $c_{A} = c_{T} =$~5.38 as shown by the dotted line. As noted above, in the low-energy limit, the phenomenological expression Eq.~(\ref{TempDep}) is less accurate and shows a quadratic dependence with $c_{T}=$~4.92. This explains why the dashed and dotted curves in Fig.~\ref{Fig6}(a,b) do not coincide at $T/T_{\rm K}<$~0.1.

It should also be possible to scale $G(B)$ as a function of a single parameter $T_{B}/T_{\rm K}$. As an example, we present in Fig.~\ref{Fig6}(a) scaled $G(B)$ data from Fig.~\ref{Fig5}(d). At low fields, the measured conductance is found to depend on $B$ according to Eq.~(\ref{QuadDep}), with the coefficient $c_{A}=c_{B}\approx c_{T}$. This equality has also been independently checked by fitting the $G(B)$  and $G(T)$ data for ${T/T_{\rm K}, T_{B}/T_{\rm K}} <$~0.1 with Eq.~(\ref{QuadDep}). The ratio between the two fit coefficients, $c_{B}/c_{T}$, is approximately 1 ($c_{B}/c_{T}=$~0.95~$\pm$~0.2), strongly counter to the theoretical expectations where $c_{B} \approx$~0.55\cite{Konik2002} and $c_{B}/c_{T} \approx$~0.101 (see Table~\ref{TableI}). To illustrate this discrepancy, we plot Eq.~(\ref{QuadDep}) with $c_{A} = c_{B} =$~0.55 in Fig.~\ref{Fig6}(a) (dash-dot). The reason for such a dramatic difference in $G(B)$ dependence between theory and experiment for both low- and intermediate-field range is unclear. We speculate that the spin-orbit interaction, previously observed in InAs nanowire-based quantum dots\cite{Fasth2007}, may play a role.

It is important to note that in order for the universal scaling $G(B)$ to be valid, the coefficient $G_{0}$ in Eqs.~(\ref{QuadDep}) and (\ref{LogDep}) should be independent of $B$. In the case of GaAs quantum dots\cite{Goldhaber-Gordon1998,Cronenwett1998,Kogan2004,Amasha2005} with $|g_{GaAs}|=$~0.44 the magnetic field required to resolve the Zeeman splitting is high and the orbital effects of that field contribute significantly, resulting in a $B$-dependent $G_{0}$, even for a field parallel to the plane of the heterostructure. In contrast, in our InAs nanowire-based quantum dot, with large $g$-factor and small dot area $S =$ 50~nm~$\times$~450~nm, Kondo resonances are suppressed (split to finite bias) at fields smaller than that required to thread one magnetic flux quantum $B<(h/e)/S\approx$~180~mT, thus making the orbital effects negligible and $G_{0}$ magnetic field independent.

Now that the scaling of the linear conductance has been established, including the stronger-than-expected effect of magnetic field, we examine how the out-of-equilibrium conductance scales as a function of bias and temperature $G/G_{0} = f(T/T_{\rm K}, eV_{sd}/k_{\rm B}T_{\rm K})$. The function used to test the universal scaling in a GaAs quantum dot,\cite{Grobis2008} and in a single-molecule device,\cite{Scott2009} originates from the low-bias expansion of the Kondo local density of states\cite{Nagaoka2002} and has the following form:
\begin{equation}
\textstyle G(T,V_{sd})=G(T,0)\left[1-\frac{c_{T}\alpha}{1+c_{T}\left(\frac{\gamma}{\alpha}-1\right)\left(\frac{T}{T_{\rm K}}\right)^{2}}\left(\frac{eV_{sd}}{k_{B}T_{\rm K}}\right)^{2}\right].
\label{VsdScaling}
\end{equation}
The coefficients $\alpha$ and $\gamma$ relate to the zero-temperature width and the temperature-broadening of the Kondo ZBA, respectively. The zero-bias conductance $G(T,0)$ is defined by Eq.~(\ref{TempQuad}). The coefficients $\alpha$ and $\gamma$ are independent of the definition of the Kondo temperature and in the low-energy limit Eq.~(\ref{VsdScaling}) reduces to the theoretically predicted expression for non-equilibrium Kondo conductance:\cite{Schiller1995}
\begin{equation}
\textstyle
\label{VsdExpansion}
    \frac{G(T,V_{sd})-G(T,0)}{c_{T}G_{0}}\approx\alpha\left(\frac{eV_{sd}}{k_{B}T_{\rm K}}\right)^{2}-c_{T}\gamma\left(\frac{T}{T_{\rm K}}\right)^{2}\left(\frac{eV_{sd}}{k_{\rm B}T_{\rm K}}\right)^{2}.
\end{equation}

The independence of $\alpha$ and $\gamma$ on the definition of Kondo temperature is important: though we have chosen an explicit definition for $T_{\rm K}$, consistent with the choice used for most quantum dot experiments and NRG calculations, other definitions may differ by a constant multiplicative factor.

Figures~\ref{Fig6}(c,d) show the scaled finite-bias conductance $(1-G(T,V_{sd})/G(T,0))/\tilde{\alpha}$, where $\tilde{\alpha} = c_{T}\alpha/(1+c_{T}(\gamma/\alpha-1))(T/T_{\rm K})^{2}$, versus $(eV_{sd}/k_{B}T_{\rm K})^{2}$, measured at different temperatures and a few values of $V_{g}$. The conductance data are fit with Eq.~\ref{VsdScaling} using a procedure described by M.~Grobis \textit{et al.}\cite{Grobis2008} with two fitting parameters $\alpha$ and $\gamma$. The range of temperatures and biases used for the fitting procedure was chosen to be close to the low-energy limit, namely $T/T_{\rm K}<0.2$ and $eV_{sd}/k_{B}T_{\rm K}\lesssim0.2$, which is comparable to the ranges used in Ref.~\onlinecite{Grobis2008}. Averaging over different points in $V_{g}$ gives $\alpha=$~0.18$\pm$0.015 and $\gamma=$~1.65$\pm$0.2 for Valley IV. Despite Valley V being in the mixed-valence regime, the parameters $\alpha$ and $\gamma$ are close to those found for Valley IV.  The scaled conductance in both cases collapses onto the same curve, shown by the dashed line, for $\pm\left(eV_{sd}/k_{\rm B}T_{\rm K}\right)^{2}\leq$~0.1, though the data from Valley V deviate more from the predicted scaling. This is not surprising because the Valley V data are in the mixed-valence regime, beside that the bias can cause additional conduction mechanisms due to proximity of the Coulomb blockade peaks.

Overall, the value of $\alpha$ obtained in our experiment is larger than previously observed in a GaAs dot\cite{Grobis2008,Yamauchi2011} ($\alpha=$~0.1) and single molecule\cite{Scott2009} ($\alpha=$~0.05). The exact reason for this discrepancy is unknown, but the smaller ratio $T_{\rm base}/T_{\rm K}$ may play a role.

There is a large number of theoretical works devoted to the universal behavior of finite-bias Kondo conductance based on both the Anderson\cite{Costi1994,Konik2002,Oguri2005,Rinc'on2009,*Rinc'on2009a,*Rinc'on2010,Roura-Bas2010,Sela2009,MunozPreprint2011,*AligiaPreprint2011} and Kondo\cite{Schiller1995,Schiller1998,Majumdar1998,Pustilnik2004,Doyon2006,Mora2009} models. \corr{Early predictions based on an exactly solvable point of the anisotropic nonequilibrium Kondo model \cite{Schiller1995,Schiller1998,Majumdar1998} yielded a value $\alpha = c_V/c_T =3/\pi^2 \approx 0.304$. This turned out to be in disagreement with experiment, which is not surprising, since this coefficient is not universal and hence will not be the same for the isotropic Kondo models. A number of subsequent papers that used a Fermi-liquid approach to treat the strong-coupling fixed point of the Kondo model\cite{Pustilnik2004,Glazman2003,*Glazman2005} or studied the $U\rightarrow\infty$ limit of the symmetric Anderson model,\cite{Oguri2005,Rinc'on2009,*Rinc'on2009a,*Rinc'on2010,Sela2009,Roura-Bas2010} all found $\alpha=3/(2\pi^{2})\approx$~0.152. Our measured
value of $\alpha = 0.18$ is in a good agreement with this prediction.  A Bethe-Ansatz treatment of the nonequilibrium Anderson model \cite{Konik2002} yielded a different result, $\alpha = 4/\pi^2$, but this was obtained using some approximations and was not claimed to be exact.

Some of the more recent theoretical papers have studied the $\alpha_V$-coefficients for the nonequilibrium Anderson model under less restrictive
conditions, i.e., allow for a left-right asymmetry and a non-infinite $U$, in an attempt to explain the  experimental results of Refs.~\onlinecite{Grobis2008,Scott2009}.} J.~Rinc\'{o}n and co-authors\cite{Rinc'on2009,*Rinc'on2009a,*Rinc'on2010} found that by setting $U$ to be finite the expected value of $\alpha$ is decreased from 0.152 to 0.1, but $\gamma$ remains $\approx$~0.5. Later, P.~Roura-Bas\cite{Roura-Bas2010} came to a similar conclusion considering the Anderson model in the strong coupling limit in both the Kondo and the mixed-valence regimes. It was shown\cite{Roura-Bas2010} that $\alpha$ reduces from 0.16 to 0.11 if some charge fluctuation is allowed by shifting from the Kondo to the mixed-valence regime, and the parameter $\gamma$ is not necessarily temperature independent. In an attempt to explain the small $\alpha$ observed in molecular devices\cite{Scott2009} Sela and Malecki\cite{Sela2009} evaluated a model for the Anderson impurity asymmetrically coupled to the leads. They concluded that deep in the Kondo regime $\alpha$ takes the value of $3/(2\pi^{2})\approx$~0.152 independent of coupling asymmetry. However, if $U$ is made finite or, in other words, some charge fluctuations are included, the parameter can vary within the range $3/(4\pi^{2}) \leq \alpha \leq 3/\pi^{2}$ (0.075~$\leq\alpha\leq$~0.3) depending on the asymmetry of the tunneling barriers. Despite the fact that our system is far from the strong coupling limit ($U\sim\Gamma$, instead of $U\gg \Gamma$, see Sec.~\ref{Kondo at zero field}), the observed value of $\alpha =$~0.18 is a good match to the strong-coupling prediction.

From temperature, magnetic field, and bias scaling of the measured conductance, we are able to define a complete set of coefficients $c_{A}$ to be used in Eq.~(\ref{QuadDep}) in order to describe the Kondo effect in the low-energy limit:
\begin{center}
    $G(T)=G_{0}[1-c_{T}\left(T/T_{\rm K}\right)^{2}]$,
    \end{center}
    \begin{center}
    $G(B)=G_{0}[1-c_{B}\left(|g|\mu_{\rm B}B/k_{\rm B}T_{\rm K}\right)^{2}]$,
    \end{center}
    \begin{center}
    $G(V_{sd})=G_{0}[1-c_{V}\left(eV_{sd}/k_{\rm B}T_{\rm K}\right)^{2}]$,
\end{center}
where $G_{0}$ is the conductance at zero temperature, magnetic field and bias, $c_{T}\approx$~5.6$\pm$1.2, $c_{B}\approx$~ 5.1$\pm$1.1, $c_{V}=c_{T}\alpha\approx$~1.01$\pm$0.27. The substantial uncertainties originate from the small number of experimental points satisfying the requirement of low temperature, field, and bias used during fitting with Eq.~(\ref{QuadDep}). Table~\ref{TableI} summarizes the experimental value of these three parameters and compares to their theoretical predictions. \corr{(The parameter $\alpha$ discussed above is denoted by $\alpha_V$ in the table.)}

\begin{table*}[!ht]
  \caption{Summary of theoretically predicted parameters $c_T$, $c_V$,
    $c_B$, and $B_c$ and their experimental values.
\corr{The second column lists
    the values of the parameters $c'_A$ appearing in $G(A) = G_0[1 -
    c'_A(A/k_{\rm B} T_0)^2]$, using a definition for the Kondo scale that
    is widespread in theoretical papers, namely, $T_0 = 1/(4 \chi_0)$, where
    $\chi_0$ is the static impurity spin susceptibility at $T=0$.
    This definition  of the Kondo temperature differs from
    the $T_{\rm K}$  used in this paper, i.e., $G(T_{\rm K})=G(0)/2$, by the factor
     $T_{\rm K}/T_0 = 0.94$.\cite{Micklitz2006}
    Thus, the coefficients $c_A$ defined in our
    Eq.~(1) and listed in the fourth column are related to those in the second by
    $c_A/c_A'= (T_{\rm K}/T_0)^2$.
    We cite only references that
    are relevant for the symmetric Anderson model in the large-$U$ limit,
    where the local occupancy is 1; generalizations for the asymmetric
    Anderson model may be found in Refs.~\onlinecite{Rinc'on2009,*Rinc'on2009a,*Rinc'on2010,Sela2009,Mora2009,MunozPreprint2011}.
    \corr{The last row lists values for the critical magnetic field $B_{c}$ beyond
    which the Kondo ZBA splits and it is expressed in units of $T_{\rm K}$ defined by Eq.~(\ref{eq:define-Tk})
    (Theory: column 2; Experiment: column 5)}.}}
\label{TableI}
\corr{\begin{ruledtabular}
\begin{tabular}{ccccl} Parameter & Predicted $c_A'$ & $\alpha_A =
c'_A/c'_T$ &$c_A = c'_A (T_{\rm K}/T_0)^2$ 
& Experimental value\\
\hline &&\\ $c_{T}$ &$\pi^4/16 \approx
6.088$\cite{Yamada1975a,*Yosida1975,*Yamada1975b,Nozi'eres1974,Costi1994,Konik2001,Konik2002,Pustilnik2004,Glazman2003,*Glazman2005}
& 1
& $5.38$
&5.6$\pm$1.2\footnote{Present experiment}
\\ $c_{V}$
& $3 \pi^2/32 \approx 0.925$\cite{Oguri2005,Rinc'on2009,*Rinc'on2009a,*Rinc'on2010,Roura-Bas2010,Sela2009,Pustilnik2004,Glazman2003,*Glazman2005,Mora2009}
& $3/(2 \pi^2) \approx 0.152$
& $0.82$
&1.01$\pm$0.27$^{\rm a}$;
0.670;\cite{Grobis2008,Yamauchi2011}
0.304\cite{Scott2009}
\\$c_{B}$
&$\pi^2/16 \approx 0.617$\cite{Konik2002,Pustilnik2004,Glazman2003,*Glazman2005}
& $1/\pi^2\approx 0.101$
& $0.55$
&5.1$\pm$1.1$^{\rm a}$ \rule[-2mm]{0mm}{0mm}
\\
\hline
$|g|\mu_{\rm B}B_{c}/k_{\rm B}T_{\rm K}$
& $1.06$\cite{Costi2000};$1.04$\cite{Hewson2005};$1.1$\cite{Wright2011}
& &
&$<$0.5$^{\rm a}$;
0.5\cite{Houck2005};
1\cite{Amasha2005};
1.5\cite{Quay2007}\rule[4.5mm]{0mm}{0mm}\\
 &&\\
\end{tabular}
\end{ruledtabular}}
\end{table*}

\begin{figure}[!ht]
  \includegraphics[width=1.00\columnwidth]{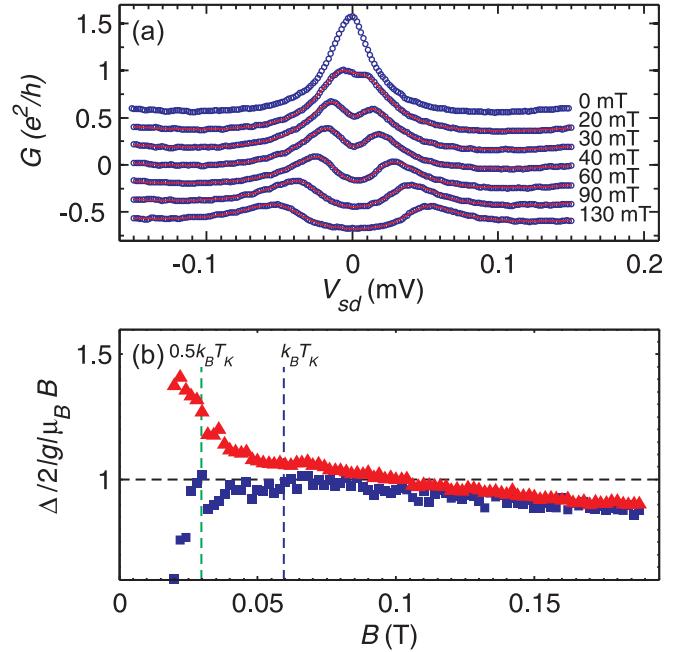}
  \caption{(a) The non-equilibrium Kondo conductance as a function of $V_{sd}$ for several values of $B$ (open blue squares). The solid red curves represent the approximation of the data made with the sum of two Fano-shaped peaks and a cubic background. (b) The normalized Zeeman splitting $\Delta/[2|g|\mu_{\rm B}B]$ as a function of $B$ data acquired from the peak maximum search (blue squares) and after fitting with two asymmetric peak shapes (red triangles). The vertical blue and green dashed lines denote magnetic field of 0.5$k_{\rm B}T_{\rm K}/|g|\mu_{\rm B}$ and $k_{\rm B}T_{\rm K}/|g|\mu_{\rm B}$ correspondingly (here $|g|=$~7.5 and $T_{\rm K}=$~300~mK).}
  \label{Fig7}
\end{figure}

\subsection{Zeeman splitting \label{Zeeman splitting}}
At non-zero magnetic field, the spin degeneracy of the Kondo singlet is lifted and the linear conductance through the dot is suppressed.\cite{Meir1993}
To recover strong transport through the dot a bias of $\pm \frac{1}{2}\Delta/e = \pm |g|\mu_{B}B/e$ should be applied in order to compensate for the spin-flip energy. As a result, in experiments, the ZBA is split into two peaks separated by $e\Delta = 2|g|\mu_{B}B/e$,\cite{Goldhaber-Gordon1998a,Cronenwett1998} providing information on the effective $g$-factor. This is why the splitting of the Kondo conductance feature has become a popular tool for evaluating the value and behavior of the $g$-factor in quantum dots made of different materials.\cite{Liang2002,Park2002,Kogan2004,Houck2005,Jespersen2006,Quay2007,Csonka2008} In this section, we discuss two unexpected features related to the Zeeman splitting. First, the minimal value of field needed to resolve the Zeeman splitting is lower than expected. Second, the splitting is weakly sublinear with magnetic field at larger fields.

Some attention has been previously paid to the value of the critical field $B_{c}$ at which the splitting of the Kondo ZBA occurs. The theory developed by one of the present authors\cite{Costi2000} predicts the value of the critical field at $T/T_{\rm K}<$~0.25 to be $B_{c} = 1.06k_{B}T_{\rm K}/|g|\mu_{\rm B}$, \corr{with similar values being found by other authors.\cite{Hewson2005,Zitko2009,Wright2011} Treating nonequilibrium more realistically gives a slightly larger value.\cite{Hewson2005} Recent work by the authors, using density matrix approaches,\cite{Hofstetter2000,Weichselbaum2007} suggests that a precise determination of the critical field is a numerically difficult task, which will require further work in order to establish this beyond any doubt. There are also somewhat conflicting experimental data on this issue.} The value of $B_{c}$ predicted by Costi\cite{Costi2000} and Hewson \textit{et al.}\cite{Hewson2005} seems to agree with the experimental findings for GaAs dots,\cite{Amasha2005} however, in gold break junctions\cite{Houck2005} the onset of the splitting was measured at 0.5$k_{B}T_{\rm K}/|g|\mu_{\rm B}$ and in the case of carbon nanotubes\cite{Quay2007} at about 1.5$k_{\rm B}T_{\rm K}/|g|\mu_{\rm B}$. In our case, $T_{\rm K} =$~300~mK (see Fig.~\ref{Fig4}(c)), \corr{thus the predicted $B_{c}$\cite{Zitko2009} is expected to be $\sim$~60~mT (for $|g|=$~7.5), more than twice as large as that observed experimentally: as seen in Fig.~\ref{Fig7}(a) and Fig.~\ref{Fig5}(c), the splitting is already well-resolved at $B=$~30~mT, which corresponds to $\sim$~0.5$k_{\rm B}T_{\rm K}/|g|\mu_{\rm B}$, the same as} the result for gold break junctions.\cite{Houck2005} Such a wide deviation of $B_{c}$ found for various Kondo systems (see Table~\ref{TableI}) may be associated with a different width of ZBA (relative to $T_K$) in the various experiments. Since the conductance peak discussed here (see Fig.~\ref{Fig4}(a)) is rather narrow, most likely due to the relatively low temperature $T/T_{\rm K}\approx$~1/30, it is possible to resolve the splitting onset at lower magnetic field. The analysis of the non-equilibrium scaling parameters, described in Sec.~\ref{Scaling subsection}, confirms the above assumption.

Finally, we discuss the evolution of the splitting $\Delta$ with magnetic field. Theory predicts that the peaks in the spectral function for spin-up and spin-down electrons should cling closer to zero energy at relatively low magnetic fields than might naively be expected, so that $\Delta$ should be suppressed by up to $\approx 1/3$ in the low-field limit.\cite{Moore2000,Konik2001,Logan2001,Costi2003,Garst2005,Weichselbaum2009} One recent experimental report corroborates this predicted trend of suppressed splitting at low field.\cite{Quay2007} But the variety of deviations from linear splitting in experiments -- especially near the onset of splitting -- is large.\cite{Amasha2005,Quay2007} To make small variations in $\Delta$ more visible, we plotted the normalized value $\delta(B) \equiv \Delta/[2|g|\mu_{\rm B}B]$ in Fig.~\ref{Fig7}(b). The value of $\Delta$ was deduced from a simple peak maximum search (blue squares) and by fitting the data with the sum of two asymmetric peak shapes and some background (red triangles). \corr{To fit $G$ as a function of $V_{sd}$ we used a combination of two Fano-shape asymmetric peaks on a cubic background:
\begin{equation}
\textstyle G(V_{sd}) = A_{1}\frac{\left[-\frac{V_{sd}+V_{1}}{\Gamma_{1}}+q_{1}\right]^2}{1+\left[-\frac{V_{sd}+V_{1}}{\Gamma_{1}}\right]^2}+A_{2}\frac{\left[\frac{V_{sd}+V_{2}}{\Gamma_{2}}+q_{2}\right]^2}{1+\left[\frac{V_{sd}+V_{2}}{\Gamma_{2}}\right]^2}+B| V_{sd} |^3+C.
\label{FittingFunction}
\end{equation}
Here $A_{1}$ and $A_{2}$ are the amplitudes, $\Gamma_{1}$ and $\Gamma_{2}$ are the widths, $q_{1}$ and $q_{2}$ are the asymmetry parameters of the two Fano resonances positioned at dc bias $V_{1}$ and $V_{2}$, respectively. Parameters $B$ and $C$ characterize the cubic conductance background. Without the cubic background, the positions of the conductance peaks, which correspond to Fano resonances at $V_{1}$ and $V_{2}$ would be $V_{p1} = V_{1}+\Gamma_{1}/q_{1}$ and $V_{p2} = V_{2}+\Gamma_{2}/q_{2}$. The peak separation is deduced from the fit according to the equation $\Delta/e=V_{p2}-V_{p1}$. The quality of this fit is shown in Fig.~\ref{Fig7}(a) by red solid curves.} It is clear that at $B>$~100~mT the splitting is sublinear in magnetic field. Coincidence of the splitting data extracted by two different methods (blue triangles and red squares in Fig.~\ref{Fig7}(b)) makes us believe that this effect is genuine and not an artifact due to weakly bias-dependent background conductance. In contrast, splitting extracted from our data at low fields $B<k_{\rm B}T_{\rm K}/|g|\mu_{\rm B}$ is dependent on the extraction method used, so we do not wish to make quantitative claims for the magnitude of splitting in that field range. Our results differ from previous observations mainly in that a sublinear field splitting occurs also at higher fields and not only at the onset of the splitting.\cite{Amasha2005,Quay2007} We are unaware of any theoretical predictions which would explain such sublinear splitting or effective reduction in the $g$-factor at higher fields.

Previous theoretical works on the Kondo model predicted a suppressed splitting $\delta(B)=\Delta/2|g|\mu_{B}B$ increasing monotonically toward one
for $g\mu_{B}B\gg k_{B}T_{\rm K}$ with logarithmic corrections.\cite{Moore2000,Rosch2003,Weichselbaum2009} For the Anderson model, similar results have been found with $\delta(B)$ rising monotonically with increasing $B$.\cite{Logan2001,Zhang2010,Zitko2011} However, in some works, \cite{Logan2001,Zhang2010,Wright2011} $\delta(B\gg k_{B}T_{\rm K})$ is found to exceed one, whereas in other works\cite{Konik2001,Zitko2011} $\delta(B\gg k_{B}T_{\rm K})$ remains below one. This discrepancy between different approaches is likely due to different approximations and the extent to which universal aspects as opposed to non-universal aspects are being addressed and remains to be clarified. For example, it is known that extracting peak positions in equilibrium spectral functions within NRG is problematic.\cite{Wright2011,Zitko2011,Schmitt2011b} Extracting a Zeeman splitting from experimental $dI/dV_{sd}$ at finite bias and large magnetic fields is also complicated by the increasing importance of higher levels and non-equilibrium charge fluctuations.\cite{Schmitt2011} Nevertheless, our results for $\delta(B\gg k_{B}T_{\rm K})$ in Fig.~\ref{Fig7}(b) exhibit a monotonically decreasing $\delta(B)$ in the high field limit for $B>1.5k_{\rm B}T_{\rm K}/|g|\mu_{\rm B}$. This contrasts to current theoretical predictions. As we cannot exclude the contribution of orbital effects at higher $B$, the magnetic fields used to determine the $g$-factor were chosen to be smaller than 100~mT (flux through dot $\le 0.6 \Phi_0$).

\section{Conclusion}
In conclusion we have performed a comprehensive study of the spin-$\frac{1}{2}$ Kondo effect in an InAs nanowire-base quantum dot. This experimental realization of a quantum dot allowed us to observe and thoroughly examine the main features of the Kondo effect including the unitary limit of conductance and dependence of the Kondo temperature on the parameters of the quantum dot. Also the Kondo temperature's quantitative relation to the Kondo ZBA shape, Zeeman splitting of the ZBA and scaling rules for equilibrium and non-equilibrium Kondo transport were studied. A previously undetected dependence of the $g$-factor on magnetic field was observed. The non-equilibrium conductance matches the previously introduced universal function of two parameters with expansion coefficients, $\alpha=$~0.18 and $\gamma=$~1.65, in quantitative agreement with predictions for the infinite-$U$ Anderson model, and consistent with the allowed range for the finite $U$ asymmetric Anderson model. We conclude that InAs nanowires are promising new objects to be used in future mesoscopic transport experiments, including highly quantitative studies.

There is one experimental observation, however, that is strikingly at odds with theoretical expectations: the conductance $G(B)$ at low temperatures shows a much stronger magnetic field dependence than expected from theoretical calculations for the single-impurity Anderson model (see Fig.~\ref{Fig5}(d)). As possible cause for this unexpected behavior, we suggest spin-orbit interactions, which are known to be strong in InAs nanowires.\cite{Fasth2007} The occurrence of a Kondo effect is compatible with the presence of spin-orbit interactions, since they do not break time-reversal symmetry. However, they will, in general, modify the nature of the spin states that participate in the Kondo effect.\cite{Sun2005,Vernek2009,Paaske2010,Pletyukhov2011} In the present geometry, where spin-orbit interactions are present in the nanowire (but not in the leads), there will be a preferred quantization direction (say $\vec n_{\rm so}$) for the doublet of local states. In general, $\vec n_{\rm so}$ is not collinear with the direction of the applied magnetic field, $\vec B$. The local doublet will be degenerate for $\vec B=0$, allowing a full-fledged Kondo effect to develop as usual in the absence of an applied magnetic field. However, the energy splitting of this doublet with increasing field will, in general, be a non-linear function of $|\vec B|$, whose precise form depends on the relative directions of $\vec B$ and $\vec n_{\rm so}$. According to this scenario, the magnetoconductance curves measured in the present work would not be universal, but would change if the direction of the applied field were varied. A detailed experimental and theoretical investigation of such effects is beyond the scope of the present paper, but would be a fruitful subject for future studies.

\begin{acknowledgments}
The authors would like to thank Moty Heiblum for making this work possible and for suggestions and critical remarks made during the work. We also acknowledge Yuval Oreg, Mike Grobis, Nancy Sandler, Sergio Ulloa, and Jens Paaske for fruitful discussions, Ronit Popovitz-Biro for the TEM analysis of nanowires, and Michael Fourmansky for technical assistance. \corr{We thank David Logan and Martin Galpin for reading an earlier version of our manuscript and pointing out that the dependence of linear conductance (but not differential conductance) on magnetic field could be understood quantitatively by taking a higher value for $g$-factor.} A.V.K. is grateful to Yunchul Chang for his design ideas and expertise in electronics. This work was partially supported by the EU FP6 Program Grant~506095, by the Israeli Science Foundation Grant~530-08 and Israeli Ministry of Science Grant~3-66799. D.G.-G. acknowledges NSF contract DMR-0906062, and U.S.-Israel BSF grant No.~2008149. \corr{T.A.C. acknowledges supercomputer support from the John von Neumann Institute for Computing (J\"{u}lich).}

\end{acknowledgments}

\bibliography{manuscript_reply}

\end{document}